\documentclass[twocolumn]{autart}    

\usepackage{graphicx}          
\usepackage[T1]{fontenc}
\usepackage[utf8]{inputenc}
\usepackage[english]{babel}
\usepackage{times}
\usepackage{enumitem}

\usepackage{fancybox,shadow,xcolor} 
\usepackage{setspace}
\usepackage{interval}
\usepackage{xcolor}

\usepackage{amssymb,amsmath}
\usepackage{mathrsfs}
\usepackage{mathdots}

\theoremstyle{definition}
\newtheorem{definition}{Definition}

\DeclareMathOperator*{\argmin}{arg\,min}

\renewcommand{\Re}{\mathbb{R}}

\usepackage{bbm}
\renewcommand{\paragraph}[1]{\smallskip\noindent\textbf{#1.} }

\newcommand{\BM}{\begin{bmatrix}}
\newcommand{\EM}{\end{bmatrix}}

\newcommand{\BBM}{\big[\begin{matrix}}
\newcommand{\EEM}{\end{matrix}\big]}

\newcommand{\bbm}{[\begin{matrix}}
\newcommand{\eem}{\end{matrix}]}

\intervalconfig {soft open fences}

\newcommand{\rev}[1]{\textcolor{black}{#1}}

\begin{document}

\begin{frontmatter}

\title{A note on the convergence of  a class of adaptive optimal identifiers\thanksref{footnoteinfo}} 

\thanks[footnoteinfo]{This paper was not presented at any IFAC 
meeting. Corresponding author L. Bako.  Tel.: +33 472 186 452}

\author[ECL]{Laurent Bako}\ead{laurent.bako@ec-lyon.fr}   

\address[ECL]{Univ Lyon, Ecole Centrale Lyon, INSA
Lyon, Universit\'{e} Claude Bernard Lyon 1,
CNRS, Amp\`{e}re, UMR 5005, 69130 Ecully,
France}  


\begin{abstract}                          
This paper  proposes a unifying framework for the convergence analysis of a class of adaptive optimal  identifiers. The considered class of identifiers is constructed from the sequence of minimizing sets of a family of objective functions. 
For the purpose of the analysis we introduce a generalized version of the classical persistence of excitation condition. Based on this an Input-to-State-Stability (ISS)-type of property is derived for the studied class of adaptive estimators.
\end{abstract}

\end{frontmatter}


\section{Introduction}
Estimation algorithms can be roughly classified into two main categories: batch methods and adaptive ones. 
While the first type of estimators produces a single estimate based on a finite collection of data, the second type generates a sequence of estimates by processing the data sequentially \cite{Goodwin09-Book}. With an appropriate weighting of the data, adaptive estimators can be endowed with the capacity of tracking slow variations in the parameter being estimated. In some cases it is even possible to find a cheap and recursive updating mechanism which computes the estimate at a time step from the one obtained at the preceding time step and the new incoming information. In these particular cases,  adaptive algorithms (e.g., the stochastic gradient descent algorithm) may present a much lower complexity than their batch counterparts, a property which is of practical interest when dealing with high dimensional or large amounts of data. 
Well-known adaptive identification algorithms include the recursive least squares (RLS with or without exponential forgetting factor) \cite{LjungBook83}, the least mean squares and their respective variants \cite{Goodwin09-Book}.

This note considers a general adaptive estimation framework for linearly parameterized systems. Its contribution consists in the unified analysis of the convergence properties of a class of optimization-based adaptive identifiers.
 Our main result is the derivation of an Input-to-State (ISS)-type \cite{Jiang2001} of property under a new generalized persistency of excitation condition. An interesting feature of the proposed analysis is that it relies (only) on the properties of the loss functions from which the estimating optimization problem is formed. That is, no explicit expression of the solution to the optimization problem is required for the analysis. This makes the approach very generic as it renders it applicable to a  variety of estimators including those resulting from smooth/nonsmooth or convex/nonconvex loss functions.  
Many existing convergence analysis methods for adaptive optimal estimators apply to a single estimator and require in general an analytic expression of the solution of the optimization problem (e.g., the analysis of the RLS in \cite{Johnstone82-SCL}). In contrast, the current paper takes a more abstract analysis  approach to  adaptive optimal identifiers thereby revealing the precise convergence-inducing properties of the minimized loss function and of the training  data.

Note that the focus of the paper is only to analyze the behavior of the sequence of estimates generated by the optimal adaptive identifier ; we do not consider implementation aspects such as computability of the estimates or numerical complexity. 

\paragraph{Notation}
$\Re$ (resp. $\Re_+$) denotes the set of real numbers (resp. nonnegative real numbers). 
We use $\Re^*$ to denote the set of nonzero real numbers and $\mathbb{N}$ for the set of natural integers. 
$\left\|\cdot\right\|$ refers to an arbitrary norm. 
$I_n$ denotes the identity matrix of dimension $n$. For two symmetric matrices $S_1$ and $S_2$ the notation $S_1\preceq S_2$  means that $S_2-S_1$ is a positive definite matrix. 
A function $\alpha:\Re_+\rightarrow\Re_+$ is said to be of class-$\mathcal{K_\infty}$ \cite{Kellett14} if it is continuous, zero at zero, strictly increasing and satisfies $\lim_{r\rightarrow +\infty }\alpha(r)=+\infty$. We denote with $\mathcal{K}_\infty$ the set of all such functions. Note that any $\alpha\in \mathcal{K}_\infty$ is invertible and its inverse, which we will denote by $\alpha^{-1}$, is also a $\mathcal{K}_\infty$ function. 

\newpage

\section{Estimation setting }
We consider a linearly parameterized data-generating system defined by
\begin{equation}\label{eq:system}
	y_t=x_t^\top \theta^\circ +v_t,
\end{equation}
where $y_t\in \Re$ is the output at  discrete time $t\in \mathbb{N}$, $x_t\in \Re^n$ is the predictor (regressor vector) and $v_t\in \Re$ represents modelling uncertainty. $\theta^\circ\in \Re^n$ is an unknown and constant parameter vector to be estimated. 

Although the following analysis is, for simplicity reason, presented here for a single output model, the results are directly applicable to multivariate measurements model (i.e., situations where the to-be-estimated parameter vector $\theta^\circ$ would be a matrix).

We consider the problem of finding a sequence $\big\{\hat{\theta}_t\big\}$ of estimates for the parameter vector $\theta^\circ$ in \eqref{eq:system} such that each estimate $\hat{\theta}_t$ relies on the data $z^t=\left\{(x_\tau,y_\tau):\tau=1,\ldots,t\right\}$ generated by the system up to time $t$. The focus of this note is to analyze the properties of a class of  optimization-based estimators which generate the sequence of estimates such that
\begin{equation}\label{eq:def-estimator}
	\hat{\theta}_t\in \argmin_{\theta\in \Re^n} V_t(\theta), 
\end{equation}
where the objective function $V_t:\Re^n\rightarrow \Re_+$ is taken to be of the form 
\begin{equation}\label{eq:Cost}
	V_t(\theta)=\sum_{k=1}^t\lambda^{t-k}\psi(y_k-x_k^\top\theta)+\lambda^t\psi_0(\theta-\hat{\theta}_0)
\end{equation}
for $t\geq 1$ and  $V_0(\theta)=\psi_0(\theta-\hat{\theta}_0)$. Here, $\hat{\theta}_0$ denotes some initial guess (or prior estimate) of the parameter vector $\theta^\circ$, \rev{$\lambda\in \interval[open]{0}{1}$} is called a forgetting factor and $\psi:\Re\rightarrow \Re_+$ and  $\psi_0:\Re^n\rightarrow \Re_+$ are some loss functions. In words, Eq. \eqref{eq:def-estimator} says that $\hat{\theta}_t$ is selected as one minimizing point of the performance index $\theta\mapsto V_t(\theta)$.

\rev{Indeed, most adaptive schemes are recursive (greedy) heuristics in that they do not necessarily minimize at each time $t$  a global objective  $V_t$ of the form \eqref{eq:Cost}. } The approach taken here is to first setup as in \eqref{eq:Cost} the cumulative performance function from which the estimates are generated. The question of whether the solution can then be  computed exactly or approximately  through a one step-ahead recursive update equation rather pertains to practical implementation aspects which are beyond the intended scope of the current paper. We rather focus on well-definedness of the sequence $\{\hat{\theta}_t\}$ from \eqref{eq:def-estimator} (i.e., existence of a minimum value for $V_t$ at each time $t$) and the convergence property.

For future reference, we consider the following properties. \\
\noindent A positive function $\ell:\Re^n\rightarrow \Re_+$ is said to 
\vspace{-4pt}
\begin{enumerate}[label=A\arabic *]  
	\item  \label{assum:positive-definite} be positive-definite if $\ell(x)=0$ $\Leftrightarrow$ $x=0$. \label{P:positive-definite}
	\item \label{assum:symmetry} be symmetric if $\ell(x)=\ell(-x)$ $\forall x\in\Re^n$.  \label{P:symmetry}
	\item \label{assum:GTI} satisfy a \textit{generalized triangle inequality (GTI)} if there exists a constant number $\alpha_{\ell}>0$ such that 
	\begin{equation}\label{eq:GTI}
	\ell(x-y)\geq \alpha_{\ell} \ell(x)-\ell(y) \quad \forall (x,y)\in \Re^n\times \Re^n. 
\end{equation}
\item \label{assum:GH} satisfy a \textit{generalized homogeneity (GH)} property if there exists a $\mathcal{K}_\infty$ function $f_{\ell}$ such that 
\begin{equation}\label{eq:GTI}
\ell(x)\geq f_{\ell}\big(\frac{1}{|r|}\big)\ell(r x)\quad \forall (x,r)\in \Re^n\times \Re^*.
\end{equation}
\end{enumerate}

\noindent {We will assume throughout the paper that the loss functions $\psi:\Re\rightarrow\Re_+$ and $\psi_0:\Re^n\rightarrow\Re_+$ are \textit{continuous} on their respective domains. Moreover, they will be required, according to the situation, to satisfy a subset of the properties \ref{assum:positive-definite}-\ref{assum:GH}. }
  Note that $\psi$ and $\psi_0$ need not be convex for the following reasoning to be valid. \rev{Examples of candidate  functions for $\psi$ may be all power functions of  the form $e\mapsto |e|^p/p$  with $p$ being a strictly positive real number, or the scalar Huber loss defined on $\Re$, for a given $h>0$, by $e\mapsto e^2/2$ if $|e|\leq h$ and $e\mapsto h(|e|-h/2)$ otherwise. } In the special case where $\psi(e)=e^2/2$ and $\psi_0(\theta)=\theta^\top P_0^{-1}\theta$ (with $P_0$ being a symmetric positive definite matrix), a closed form expression of the solution $\hat{\theta}_t$ can indeed be obtained; moreover, that solution can be written in a recursive form (called the RLS algorithm, see, e.g., \cite{LjungBook83}).
 
\begin{lem}\label{lem:bounds-loss}
Let $\ell:\Re^{n}\rightarrow\Re_{+}$ be a continuous function which has the properties \ref{assum:positive-definite}--\ref{assum:symmetry} and~\ref{assum:GH} with a $\mathcal{K}_\infty$ function $f_{\ell}$. Then, for all norm $\lVert \cdot \rVert$ on  $\Re^{n}$, there exist $\xi_1,\xi_2\in \mathcal{K}_{\infty}$ such that
\begin{equation}\label{eq:fl_r}
\xi_1(\left\|\theta\right\|)\leq \ell(\theta) \leq \xi_2(\left\|\theta\right\|) \quad \forall \theta\in \Re^{n}. 
\end{equation}
A particular choice of $\xi_1$ and $\xi_2$ is given by $\xi_1(r)=D_1 f_\ell(r)$ and $\xi_2(r)= D_2\dfrac{1}{f_\ell(1/r)} $ for $r\neq 0$ and $\xi_2(0)=0$, with $D_1=\min_{\lVert \theta \rVert=1} \ell(\theta)>0$ and  $D_2=\max_{\lVert \theta \rVert=1} \ell(\theta)>0$.
\end{lem}
\begin{pf}
The proof follows similar arguments as that of Lemma 2  in \cite{Kircher21-TAC}. First, by the continuity of $\ell$ we know that it admits minimum value $D_1$ and maximum value $D_2$ on the compact set $\mathcal{S}=\left\{\theta\in \Re^n:\left\|\theta\right\|=1\right\}$. Second, by the generalized homogenous property \ref{assum:GH} of $\ell$ we have  
\begin{equation}\label{eq:Ineq-GH}
	f_\ell(\left\|\theta\right\|)\ell\left(\frac{\theta}{\left\|\theta\right\|}\right)\leq \ell(\theta)\leq \dfrac{1}{f_\ell(1/\left\|\theta\right\|)} \ell\left(\frac{\theta}{\left\|\theta\right\|}\right)
\end{equation}
for all $\theta\neq 0$. The result follows by noting that $\theta/\left\|\theta\right\|$ lies in $\mathcal{S}$, which implies that 
$D_1\leq \ell(\theta/\left\|\theta\right\|)\leq D_2$. \qed
\end{pf}

\section{General convergence analysis}

We start the analysis by introducing a concept of persistence of excitation on the predictor sequences $\left\{x_t\right\}$. 
\begin{definition}\label{def:PE}
The sequence $\left\{x_t\right\}$ is said to be persistently exciting (PE) with respect to the loss function $\psi:\Re\rightarrow\Re_+$ (subject to \ref{assum:positive-definite}--\ref{assum:symmetry} and~\ref{assum:GH}) if there exist two $\mathcal{K}_\infty$ functions $\alpha$ and $\beta$ and a fixed time horizon $T>0$ such that 
\begin{equation}\label{eq:PE}
	\alpha(\left\|\theta\right\|)\leq \sum_{k=t+1}^{t+T}\psi(x_k^\top \theta)\leq \beta(\left\|\theta\right\|) \quad \forall (t,\theta)\in \mathbb{N}\times \Re^n
\end{equation}
for some norm $\left\|\cdot\right\|$ on $\Re^n$. 
\end{definition}
\noindent Definition \ref{def:PE} can be interpreted as a generalization of the classical PE condition, see e.g., \cite{Johnstone82-SCL}. 
\rev{For example,  in the special case where $\psi$  and the $\mathcal{K}_\infty$ functions $\alpha$ and  $\beta$ are defined respectively by  $\psi(e)=e^2$, $\alpha(r)=ar^2$ and $\beta(r)=br^2$ for some $a>0$ and $b>0$, Eq.\eqref{eq:PE} becomes, when Euclidean norm is substituted for the generic norm $\left\|\cdot\right\|$,  
$$	a \theta^\top \theta\leq \sum_{k=t+1}^{t+T}\theta^\top  x_k x_k^\top \theta\leq b \theta^\top \theta \quad \forall (t,\theta)\in \mathbb{N}\times \Re^n. $$
The latter is equivalent to 
$aI_n\preceq \sum_{k=t+1}^{t+T}x_kx_k^\top\preceq  bI_n \quad \forall t\in \mathbb{N}.$
The latter corresponds to the classical PE condition \cite{Johnstone82-SCL}.
} The PE condition \eqref{eq:PE} can indeed be equivalently reformulated as follows. 

\begin{lem}
Assume that $\psi$ is continuous and satisfies \ref{assum:positive-definite}-\ref{assum:symmetry} and \ref{assum:GH}. Then Eq. \eqref{eq:PE}  holds true if and only if there exist positive constants $\gamma_1$ and $\gamma_2$ such that
\begin{equation}\label{eq:SC}
	\gamma_1\leq \sum_{k=t+1}^{t+T}\psi\big(\frac{x_k^\top \theta}{\|\theta\|}\big)\leq \gamma_2 \quad \forall (t,\theta)\in \mathbb{N}\times \Re^n\setminus\left\{0\right\}. 
\end{equation}
\end{lem}
\vspace{-1cm}
\begin{pf}
The proof follows the same idea as that of Lemma \ref{lem:bounds-loss}. In effect, if \eqref{eq:SC} holds then by applying the  inequality  \eqref{eq:Ineq-GH} with $\ell(\theta)=\sum_{k=t+1}^{t+T}\psi\big(x_k^\top \theta\big)$ we immediately see that \eqref{eq:PE} holds with $\mathcal{K}_{\infty}$ functions $\alpha$ and $\beta$  defined respectively by $\alpha(r)=\gamma_1 f_\psi(r)$ and $\beta(r)= \gamma_2  \frac{1}{f_\psi(1/r)}$ for $r\neq 0$ and $\beta(0)=0$. 

Conversely, to see that \eqref{eq:PE} implies \eqref{eq:SC}, it suffices to replace $\theta$ with $\theta/\left\|\theta\right\|$ in \eqref{eq:PE} and take $\gamma_1=\alpha(1)$ and $\gamma_2=\beta(1)$. 
\qed
\end{pf}
\begin{lem}\label{lem:positive-definite}
Let $\psi:\Re\rightarrow \Re_+$ and $\psi_0:\Re^n\rightarrow \Re_+$  be continuous  functions satisfying the properties \ref{assum:positive-definite}-\ref{assum:symmetry} and \ref{assum:GH}. Consider a function $G_t:\Re^n\rightarrow \Re_+$ defined by 
\begin{equation}\label{eq:Gt}
	G_t(\theta)= \alpha_{\psi}\sum_{k=1}^t\lambda^{t-k}\psi(x_k^\top \theta)+\alpha_{\psi_0}\lambda^t\psi_0(\theta), 
\end{equation}
where $x_k\in\Re^n$ $\forall k$, $\alpha_{\psi}>0$, $\alpha_{\psi_0}>0$ and \rev{$\lambda\in \interval[open]{0}{1}$}. 
If $\left\{x_t\right\}$ satisfies the PE condition \eqref{eq:PE} with respect to the function $\psi$,  then there exist  time-invariant $\mathcal{K}_\infty$ functions $g_1$ and $g_2$ such that 
\begin{equation}\label{eq:bounds-Gt}
	g_1(\left\|\theta\right\|) \leq G_t(\theta)\leq g_2(\left\|\theta\right\|) \quad \forall (t,\theta)\in \mathbb{N}\times \Re^n. 
\end{equation}
\end{lem}
\begin{pf}
Since $\left\{x_t\right\}$ is PE with respect to $\psi$, there exist $\alpha\in\mathcal{K}_\infty$ and a number $T>0$ such that \eqref{eq:PE} holds. For any $t\in \mathbb{N}$, consider the pair $(q_t,r_t)$ of integers denoting the quotient and the rest of the euclidean division of $t$ by $T$, i.e., such that  $t=q_tT+r_t$ with $0\leq r_t<T$. \\
Then by a similar reasoning as in \cite{Bako16-Automatica-b}, the following series of inequalities holds for $t\geq T$: 
$$
	\begin{aligned}
	G_t(\theta) &\geq \lambda^t \sum_{i=1}^{q_t} \sum_{k=(i-1)T+1}^{iT}\lambda^{-k}\alpha_{\psi}\psi(x_k^\top\theta\big)\\
	&= \lambda^{(q_t+1)T-1} \sum_{i=1}^{q_t} \lambda^{-(i-1)T} \sum_{k=(i-1)T+1}^{iT}\alpha_{\psi}\psi(x_k^\top\theta\big).
\end{aligned}
$$
Recalling that $ \sum_{k=(i-1)T+1}^{iT}\psi(x_k^\top\theta\big)\geq  \alpha(\left\|\theta\right\|)$ by the PE condition, we can write
$$
\begin{aligned}	
	&G_t(\theta)\geq  \lambda^{(q_t+1)T-1} \sum_{i=1}^{q_t} \lambda^{-(i-1)T} \alpha_{\psi}\alpha(\left\|\theta\right\|)\\
	& = \lambda^{(q_t+1)T-1}\dfrac{1-\lambda^{-q_tT}}{1-\lambda^{-T}}\alpha_{\psi}\alpha(\left\|\theta\right\|)\geq g_{11}(\left\|\theta\right\|)
\end{aligned}
$$
with $  g_{11}(r) = \alpha_{\psi}\lambda^{2T-1}\alpha(r)$. 
On the other hand, for $t<T$, we can write $G_t(\theta)\geq \alpha_{\psi_0}\lambda^{T-1}\psi_0(\theta)$. Since $\psi_0$ is subject to the conditions \ref{assum:positive-definite}-\ref{assum:symmetry} and \ref{assum:GH}, we conclude from Lemma \ref{lem:bounds-loss} that there exists  $g_{12}\in \mathcal{K}_{\infty}$  such that $\alpha_{\psi_0}\lambda^{T-1}\psi_0(\theta)\geq g_{12}(\left\|\theta\right\|)$ for $0\leq t<T$. 
Taking then $g_1=\min(g_{11},g_{12})$, it can be shown that $g_1\in \mathcal{K}_\infty$. Hence,  the lower bound in \eqref{eq:bounds-Gt} is proved. 

\noindent We now turn to proving the upper bound. For all $t\geq T$:  
$$\begin{aligned}
	G_t(\theta)&\leq \lambda^t\sum_{i=1}^{q_t+1} \sum_{k=(i-1)T+1}^{iT}\lambda^{-k}\alpha_{\psi}\psi(x_k^\top \theta)+\alpha_{\psi_0}\lambda^T\psi_0(\theta)\\
	&\leq \alpha_{\psi}\lambda^T\sum_{i=1}^{q_t+1}\lambda^{-iT} \sum_{k=(i-1)T+1}^{iT}\psi(x_k^\top \theta)+\alpha_{\psi_0}\lambda^T\psi_0(\theta)\\
	&\leq \alpha_{\psi}\lambda^T\sum_{i=1}^{q_t+1} \lambda^{-iT}\beta(\left\|\theta\right\|)+\alpha_{\psi_0}\lambda^T\psi_0(\theta)\\
	&\leq \alpha_{\psi}\dfrac{1-\lambda^{-2T}}{1-\lambda^{-T}}\beta(\left\|\theta\right\|)+\alpha_{\psi_0}\lambda^T\psi_0(\theta).
\end{aligned} 
$$
The right hand side function satisfies the conditions of Lemma \ref{lem:bounds-loss}. 
Therefore, we can find a time-invariant $g'\in \mathcal{K}_\infty$ such that $G_t(\theta)\leq g'(\left\|\theta\right\|)$. 
Consider now the case where $0\leq t<T$. For each such $t$, $G_t(\theta)$ fulfills the conditions of Lemma \ref{lem:bounds-loss}. As a consequence, there exist $\mathcal{K}_\infty$ functions $g_{2,t}$, $t=0,\ldots,T-1$, such that
$G_t(\theta)\leq g_{2,t}(\left\|\theta\right\|) $. It suffices then to take $g_2=\max(g',g_{2,0},\ldots,g_{2,T-1})$ (which is in $\mathcal{K}_{\infty}$) for the upper bound in  \eqref{eq:bounds-Gt} to hold. 
\qed
\end{pf}
\vspace{-.5cm}
The next lemma gives sufficient conditions for the the estimates sequence $\{\hat{\theta}_t\}$ to be well-defined. 
\begin{lem}
Consider the performance index \eqref{eq:Cost} with $\psi$ and $\psi_0$ satisfying all the properties  \ref{assum:positive-definite}--\ref{assum:GH} along with continuity. If the regressor sequence $\left\{x(t)\right\}$ satisfies the PE condition stated in \eqref{def:PE}, then the sequence of estimates \eqref{eq:def-estimator} is well-defined, that is, $\argmin_{\theta} V_t(\theta)\neq \emptyset$ for all $t\in \mathbb{N}$. 
\end{lem}
\begin{pf}
The idea of the proof is to show that $V_t$ is coercive (i.e., continuous and radially unbounded) and then apply a result in \cite{Rockafellar05-Book} to conclude on the existence of a minimizer of $V_t$ for all $t$. For this purpose, note that $V_t$ is continuous as a consequence of $\psi$ and $\psi_0$ being continuous.  Moreover,  from the properties \ref{assum:symmetry}-\ref{assum:GTI} of $\psi$ and $\psi_0$, one can write 
$$
\begin{aligned}
			V_t(\theta)&\geq  G_t(\theta)-\sum_{k=1}^t\lambda^{t-k}\psi(y_k)-\lambda^t\psi_0(\hat{\theta}_0)\\
			&\geq g_1(\left\|\theta\right\|)-\sum_{k=1}^t\lambda^{t-k}\psi(y_k)-\lambda^t\psi_0(\hat{\theta}_0)
\end{aligned}
$$
for some norm $\left\|\cdot\right\|$ on $\Re^n$. Here, $g_1$ is defined as in \eqref{eq:bounds-Gt}.  From the inequality above it is clear that $\lim_{\left\|\theta\right\|\rightarrow \infty} V_t(\theta)=+\infty$.  
Hence, we conclude that $V_t$ is coercive. As a consequence, $V_t$ admits  a minimum value for all $t$. The sequence of estimates \eqref{eq:def-estimator} is therefore well-defined as claimed. \qed
\end{pf}
The claim of the lemma is that $V_t$ admits a minimum value but  the minimizer of $V_t(\theta)$ need not be unique. The potential nonuniqueness of the sequence $\big\{\hat{\theta}_t\big\}$ is not an issue as the derived convergence property holds for any sequence of estimate satisfying  \eqref{eq:def-estimator}.  
\\
\begin{thm}\label{thm:cvgce}
Consider a parameter sequence $\big\{\hat{\theta}_t\big\}$ generated as in \eqref{eq:def-estimator} with the assumption that the functions $\psi$ and $\psi_0$ in \eqref{eq:Cost} are continuous and satisfy the conditions \ref{assum:positive-definite}--\ref{assum:GTI}. If the sequence $\left\{x_t\right\}$ of regressors satisfies the PE condition as stated in Definition \ref{def:PE}  for a given norm $\left\|\cdot\right\|$ on $\Re^n$, then there exists  $\xi \in \mathcal{K}_\infty$  such that 
\begin{equation}\label{eq:ISS}
	\|\hat{\theta}_t-\theta^\circ\|\leq \xi^{-1} \Big[2\lambda^t\psi_0(\hat{\theta}_0-\theta^\circ)+2\sum_{k=1}^{t}\lambda^{t-k}\psi(v_{k})\Big]
\end{equation}
for all $t\in \mathbb{N}$. 
\end{thm}
\begin{pf}
By invoking Eq. \eqref{eq:system}, it is easy to see that 
$$V_t(\theta)= \sum_{k=1}^t\lambda^{t-k}\psi(v_k-x_k^\top(\theta-\theta^\circ))+\lambda^t\psi_0((\theta-\theta^\circ)-(\hat{\theta}_0-\theta^\circ)).$$
Now pose $\eta_t=\hat{\theta}_t-\theta^\circ$ so that $\eta_0=\hat{\theta}_0-\theta^\circ$.  Then we can observe that
$$
\begin{aligned}
	&V_t(\hat{\theta}_t)= \sum_{k=1}^t\lambda^{t-k}\psi(v_k-x_k^\top\eta_t)+\lambda^t\psi_0(\eta_t-\eta_0)
\end{aligned}
$$
and
$$
V_t(\theta^\circ)=\sum_{k=1}^t\lambda^{t-k}\psi(v_k)+\lambda^t\psi_0(\eta_0). 
$$
By the generalized triangle inequality property  of $\psi$ and $\psi_0$, we have $\psi(v_k-x_k^\top\eta_t)\geq \alpha_{\psi}\psi(x_k^\top \eta_t)-\psi(v_k)$ and $\psi_0(\eta_t-\eta_0)\geq \alpha_{\psi_0}\psi_0(\eta_t)-\psi_0(\eta_0)$. Making use of these inequalities give
$$G_t(\eta_t)-V_t(\theta^\circ)\leq V_t(\hat{\theta}_t)\leq V_t(\theta^\circ), $$
where $G_t$ is defined as in \eqref{eq:Gt}. In the above equation, the right hand side inequality follows from the definition of $\hat{\theta}_t$ as a minimizer of $V_t$.  
Therefore, we have  
$G_t(\eta_t)\leq 2V_t(\theta^\circ). $
By applying Lemma \ref{lem:positive-definite} for a given norm $\left\|\cdot \right\|$ on $\Re^n$, there exists a time-invariant $\mathcal{K}_\infty$ function $\xi$ such that $G_t(\theta)\geq \xi(\left\|\theta\right\|)$. 
Combining this with the inequality above yields
$$ \left\|\eta_t\right\| \leq \xi^{-1}\left(2V_t(\theta^\circ)\right)$$
which is the stated result stated by the theorem. 
 \qed
\end{pf}
\rev{Theorem \ref{thm:cvgce} states a property of ISS (see, e.g., \cite{Jiang2001} for a definition) for the system $\big(\eta_0,\left\{v_t\right\}\big)\mapsto \left\{\eta_t\right\}$ which maps the initial parametric error $\eta_0=\hat{\theta}_0-\theta^\circ$ and the measurement noise $\left\{v_t\right\}$ to the  estimation error sequence $\left\{\eta_t\right\}$. What it says is that the estimator forgets exponentially fast the initial  condition and produces an error which is bounded by an increasing function of the amount of  noise.}

Assuming that the noise sequence $\left\{v_t\right\}$ is bounded, we obtain an immediate consequence of Theorem \ref{thm:cvgce} \rev{in the form of an asymptotic upper bound on the estimation error.} 
\begin{cor}
In addition to the conditions of  Theorem \ref{thm:cvgce}, assume that the noise sequence  $\left\{v_t\right\}$ is bounded and \rev{$\lambda\in \interval[open]{0}{1}$}. Then there exists a $\mathcal{K}_\infty$ function $\xi$ such that  
$$\limsup_{t\rightarrow +\infty} \big\|\hat{\theta}_t-\theta^\circ\big\|\leq \xi^{-1}\left(\dfrac{2}{1-\lambda}\sup_{t\in \mathbb{N}}\psi(v_t)\right) $$
for any sequence $\big\{\hat{\theta}_t\big\}$ generated according to \eqref{eq:def-estimator}. 
\end{cor}
A direct implication of this corollary is that in the noise-free case, we have $\lim_{t\rightarrow+\infty}\hat{\theta}_t=\theta^\circ$. More, it can be seen from Eq. \eqref{eq:ISS} that the associated rate of convergence is exponential.

\paragraph{Example 1}
If $\psi$ and $\psi_0$ are instantiated respectively by  $\psi(e)=|e|^p$ for a real $p>0$ and $\psi_0(\theta)=\gamma_0 \left\|\theta\right\|^2$, then the properties  \ref{assum:positive-definite}-\ref{assum:GTI} are satisfied. In particular, the constant $\alpha_\psi$ in \eqref{eq:GTI} can be taken equal to $2^{1-1/p}$ if $0<p\leq 1$ and $2^{1-p}$ otherwise. Moreover, we may take the functions $\alpha$ and $\beta$ in the PE condition \eqref{eq:PE} to be of the form $\alpha(r)=ar^p$ and $\beta(r)=br^p$ respectively for some positive constants $a$ and $b$.   Using now  the proofs of Lemma \ref{lem:positive-definite} and Theorem \ref{thm:cvgce} we determine a valid candidate for the  $\mathcal{K}_\infty$ function $\xi$ in \eqref{eq:ISS} as $\xi(r)=\min\left( a\alpha_\psi\lambda^{2T-1}r^p, \gamma_0/2\lambda^{T-1}r^2\right)$.

\section{Conclusion}
This paper has proposed a unified framework for convergence analysis of a class of adaptive optimization-based estimators. More precisely, an ISS-type of property has been derived for the parametric estimation error under a generalized condition of persistence of excitation. 
An interesting feature of the proposed analysis is that it relies only on the assumed properties of the loss function entering the definition of the to-be-minimized objective function, that is, no closed-form expression of the estimates is required in the analysis. 
\rev{This makes the analysis applicable to a wide range of optimal adaptive estimators. Moreover, it  characterizes the precise convergence-inducing properties of the objective function and of the training data. 
} 
\vspace{-.3cm}
\bibliographystyle{abbrv}

\end{document}